\documentclass[%
 aip,
 amsmath,amssymb,
 reprint,%
]{revtex4-1}
\usepackage{textgreek} % text greek letters
\usepackage{graphicx}% Include figure files
\usepackage{dcolumn}% Align table columns on decimal point
\usepackage{bm}% bold math
\usepackage[utf8]{inputenc}
\usepackage[T1]{fontenc}
\usepackage{mathptmx}
\usepackage{etoolbox}
\newcommand{\minus}{\scalebox{0.75}[1.0]{$-$}}

\makeatletter
\def\@email#1#2{%
 \endgroup
 \patchcmd{\titleblock@produce}
  {\frontmatter@RRAPformat}
  {\frontmatter@RRAPformat{\produce@RRAP{*#1\href{mailto:#2}{#2}}}\frontmatter@RRAPformat}
  {}{}
}%
\makeatother
\begin{document}

\preprint{AIP/123-QED}

\title[]{Optimization of Superconducting Niobium Nitride Thin Films\\via High-Power Impulse Magnetron Sputtering}

\author{Hudson T. Horne}
\affiliation{$^{1)}$Department of Physics, University of North Florida, Jacksonville, Florida 32224 USA}

\author{Collin M. Hugo}
\affiliation{$^{1)}$Department of Physics, University of North Florida, Jacksonville, Florida 32224 USA}

\author{Brandon C. Reid}
\affiliation{$^{1)}$Department of Physics, University of North Florida, Jacksonville, Florida 32224 USA}

\author{Daniel F. Santavicca}
\affiliation{$^{1)}$Department of Physics, University of North Florida, Jacksonville, Florida 32224 USA}
\email{daniel.santavicca@unf.edu}

\date{\today}

\begin{abstract}
We report a systematic comparison of niobium nitride thin films deposited on oxidized silicon substrates by reactive DC magnetron sputtering and reactive high-power impulse magnetron sputtering (HiPIMS). After determining the nitrogen gas concentration that produces the highest superconducting critical temperature for each process, we characterize the dependence of the critical temperature on film thickness. The optimal nitrogen concentration is higher for HiPIMS than for DC sputtering, and HiPIMS produces higher critical temperatures for all thicknesses studied. We attribute this to the HiPIMS process enabling the films to get closer to optimal stoichiometry before beginning to form a hexagonal crystal phase that reduces the critical temperature, along with the extra kinetic energy in the HiPIMS process improving crystallinity. We also study the ability to increase the critical temperature of the HiPIMS films through the use of an aluminum nitride buffer layer and substrate heating.
\end{abstract}

\maketitle

\section{Introduction}
Niobium nitride (NbN) is a refractory metal with a superconducting critical temperature ($T_\mathrm{C}$) as high as $\approx 16$ K.\cite{shy} NbN thin films are utilized in a wide range of superconducting device applications, including photodetectors such as hot electron bolometers\cite{gousev} and superconducting nanowire single-photon detectors;\cite{holzman} and Josephson junction devices such as SIS heterodyne mixers,\cite{mcgrath} RSFQ digital logic,\cite{shao} and superconducting qubits.\cite{kim} Due to their high kinetic inductivity,\cite{Annunziata} NbN thin films are of interest in applications such as slow-wave transmission lines,\cite{santavicca} tunable couplers,\cite{colangelo} and travelling wave parametric amplifiers.\cite{adamyan} For many of these applications, very thin films ($\lesssim 10$ nm) are required, and hence a number of groups have explored different techniques for producing high-quality ultra-thin NbN films.\cite{ilin2008ultra,espiau2007microstructure,iovan,chockalingam,wright,schneider2009structural,shiino,rhazi2021improvement} 

Several crystal phases of NbN can occur under normal conditions. The cubic NaCl-like $\delta$-phase has the highest $T_\mathrm{C}$ and occurs when a near-stoichiometric reaction between nitrogen and niobium takes place.\cite{shy} At a $1{:}1$ stoichiometry, however, the most stable crystal phase is the hexagonal $\epsilon$-phase, which has a lower $T_\mathrm{C}$. On the other hand, reducing the nitrogen concentration can lead to the formation of tetragonal Nb$_4$N$_3$ or hexagonal Nb$_2$N, which also have a lower $T_\mathrm{C}$.\cite{benkahoul2004structural, wright} Optimizing the nitrogen concentration is thus important for maximizing the $T_\mathrm{C}$ of NbN thin films.  

Other factors also affect the film structure and $T_\mathrm{C}$, including the choice of substrate and its temperature during deposition. While Si substrates are often desired due to their availability in larger wafer sizes and their compatibility with standard fabrication processes, MgO and sapphire substrates have been shown to produce NbN films with higher $T_\mathrm{C}$ due to their improved lattice match with NbN.\cite{miki2007nbn, kang2003fabrication, espiau2007microstructure}
The use of an AlN buffer layer has also been shown to improve $T_\mathrm{C}$ on substrates such as Si.\cite{shiino, rhazi2021improvement} 

Heating the substrate can improve the crystallinity of the film, and single-domain epitaxial films can be produced using an appropriate substrate at sufficiently high temperature.\cite{kim,wright,iovan,chockalingam} These high temperatures, however, are not compatible with some materials and fabrication processes, including lift-off processes and standard CMOS processes. Single-domain epitaxial films also have a lower normal-state resistivity and hence a lower kinetic inductivity than polycrystalline films,\cite{Annunziata} and as a result they may not be desirable for applications that utilize the high kinetic inductivity of NbN. 

Sputter deposition is the most widely used approach for producing NbN thin films. High-power impulse magnetron sputtering (HiPIMS) is a sputter deposition technique in which a DC power supply is operated in a pulsed mode, resulting in high-intensity plasma excitation during the ${\sim100}$ $\mu$s pulses.\cite{lundin, gaines} This translates to a higher kinetic energy of the sputtered target material relative to conventional DC sputtering. Recent work found an increased $T_\mathrm{C}$ in NbN films deposited via HiPIMS as compared to DC sputtering.\cite{kalal2022study} This work studied relatively thick films deposited on a Si substrate with a TiN buffer layer. Here, we report a systematic comparison of DC sputtered and HiPIMS films deposited on oxidized silicon substrates, including the dependence of $T_\mathrm{C}$ on the nitrogen gas concentration during deposition and the film thickness. We also study the use of an AlN buffer layer and substrate heating with the HiPIMS process.   

\section{Experimental Parameters}
All depositions were performed in a Kurt J. Lesker PRO Line PVD 75 system with $3$-inch diameter sputter targets. The system is pumped by a turbomolecular pump backed by an oil-free scroll pump and the chamber has a base pressure of $<1 \times 10^{-7}$ torr. The system uses a sputter-up configuration with a vertical distance of approximately $12$ cm between the targets and the sample stage. The sample stage was rotated at $10$ rpm during all depositions. The sample stage can be heated to $350$ $^{\circ}$C using a quartz lamp heater.

NbN films were deposited using a Nb target in a reactive process with nitrogen gas. The Nb target was specified as 99.95\% pure (excluding tantalum). Ultra high purity argon gas (99.9999\% purity) and ultra high purity nitrogen gas (99.999\% purity) were introduced into the chamber via separate mass flow controllers. The chamber was pumped to $\lesssim 5 \times 10^{-7}$ torr prior to deposition. During all depositions, a gate valve is used to throttle the turbomolecular pump and the mass flow controllers set the gas flow rates to achieve a chamber pressure of $3.0 \times 10^{-3}$ torr. The nitrogen gas concentration is specified as a percentage of the argon gas flow rate (not as a percentage of the total gas flow). For example, an argon flow rate of $12.5$ sccm and a nitrogen flow rate of $2.5$ sccm would be specified as a $20$\% nitrogen gas concentration.

\begin{figure}
\includegraphics[width=8.5cm]{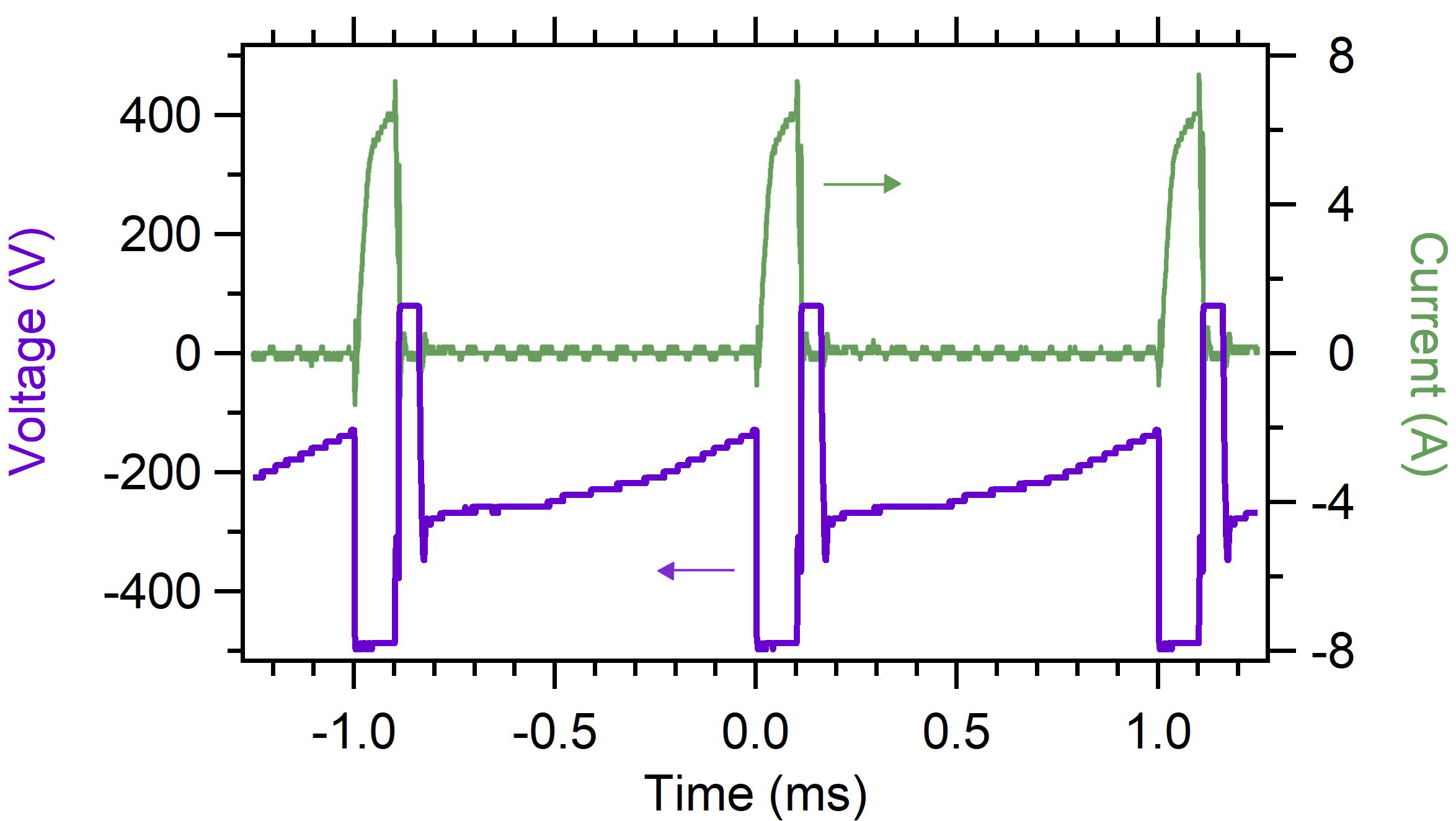}
\caption{\label{fig:Pulses} Voltage and current pulses used in the HiPIMS process. The pulses correspond to a time-average power of $250$ W and a peak power of approximately $3$ kW.}
\end{figure}

Depositions were performed using both DC sputtering and HiPIMS. The HiPIMS depositions utilized a unipolar DC power supply connected to an Impulse $2$ kW pulsed power module from Starfire Industries. This module allows for a user-configurable voltage pulse height, width, and repetition rate. All HiPIMS depositions described here used a $1$ kHz repetition rate and a $100$ $\mu$s pulse width. The HiPIMS was performed in a power-limited mode set to a time-average power of $250$ W; this determined the resulting voltage pulse height, which had a value of approximately $\minus 480$ V. The main pulse was followed by a positive kick pulse with an amplitude of approximately $+50$ V and a duration of $50$ $\mu$s. This kick pulse has been shown to increase the deposition rate by driving ionized target material to the sample, as the high peak power of the HiPIMS process can result in a higher ionization fraction of the target material.\cite{gaines, kalal2022study} The kick pulse has also been shown to increase the average ion kinetic energy.\cite{BipolarHiPIMS} An example of the resulting voltage and current pulses is shown in figure \ref{fig:Pulses}. The peak power, which occurs at the maximum of the current pulse, is approximately $3$ kW.

The NbN films deposited with DC sputtering used a power of $150$ W. The different time-average powers for the DC and HiPIMS depositions were chosen to achieve a similar deposition rate of $\approx 1$ {\AA}/s, although the deposition rate decreased as the concentration of nitrogen gas was increased, as discussed in the next section.

Unless otherwise specified, films were grown on high-resistivity Si substrates ($\rho > 5$ k$\Omega$ cm) with a native oxide layer. For some films, an aluminum nitride (AlN) buffer layer was deposited on the substrate before the NbN deposition. This was accomplished using a reactive process in which an aluminum target ($99.99$\% purity) was sputtered with a $30$\% nitrogen gas concentration.\cite{shiino,rhazi2021improvement} This deposition was performed using a DC power supply at $100$ W, corresponding to a voltage of $\minus 248$ V and a current of $0.40$ A, and resulting in a deposition rate of $0.31$ {\AA}/s. The AlN film was confirmed to be electrically insulating at room temperature, and its complex index of refraction, measured with a spectroscopic ellipsometer from $380-900$ nm, was consistent with established values.\cite{Adachi}

Electrical measurements of the NbN films were performed in a dipper-style liquid helium cryostat with the samples inside a vacuum can. The system has a $1$ K pot and a base temperature of approximately $1.5$ K. The films were mounted directly onto a copper sample holder using GE varnish, and the copper sample holder was attached to a copper block connected to the $1$ K pot. A thermometer and resistive heater are also mounted to the copper block. Ten Manganin wires (36 AWG) connect between the sample space and room temperature, with each pair of wires having a 2-wire lead resistance of $120$ $\Omega$. Aluminum wirebonds provide the electrical connection between the film and sample holder. The device resistance is measured with a lock-in amplifier using an AC excitation current with an RMS amplitude of $1$ $\mu$A, and resistance versus temperature curves are taken by first cooling the device below $T_\mathrm{C}$, then using the heater to heat the sample up to $18$ K, and then monitoring the device voltage as the heater voltage ramps down to gradually cool the sample through its superconducting transition. The rate of temperature decrease was approximately $0.3$ K per minute, ensuring negligible thermal lag between the sample and thermometer.  

\section{Comparison of DC and HiPIMS Films}

\begin{figure}
\includegraphics[width=8cm]{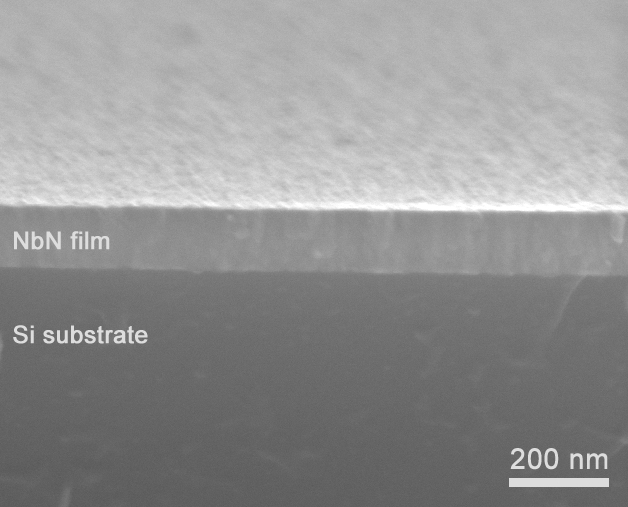}
\caption{\label{fig:SEM} Scanning electron micrograph of cross-section of 120 nm thick NbN film deposited via HiPIMS at $10\%$ nitrogen gas concentration.}
\end{figure}

NbN films were deposited onto unheated substrates at varying nitrogen gas concentrations using both DC sputtering and HiPIMS. The deposition rate varies with the nitrogen gas concentration, so the deposition rate had to be determined for each nitrogen concentration. This was done by performing a $900$ s deposition, scribing and snapping the resulting sample, and imaging the sample's cross-section in a scanning electron microscope (SEM) to determine the film thickness. An example SEM image of a $120$ nm thick film deposited via HiPIMS at $10\%$ nitrogen gas concentration is shown in figure \ref{fig:SEM}.

\begin{figure}
\includegraphics[width=8.5cm]{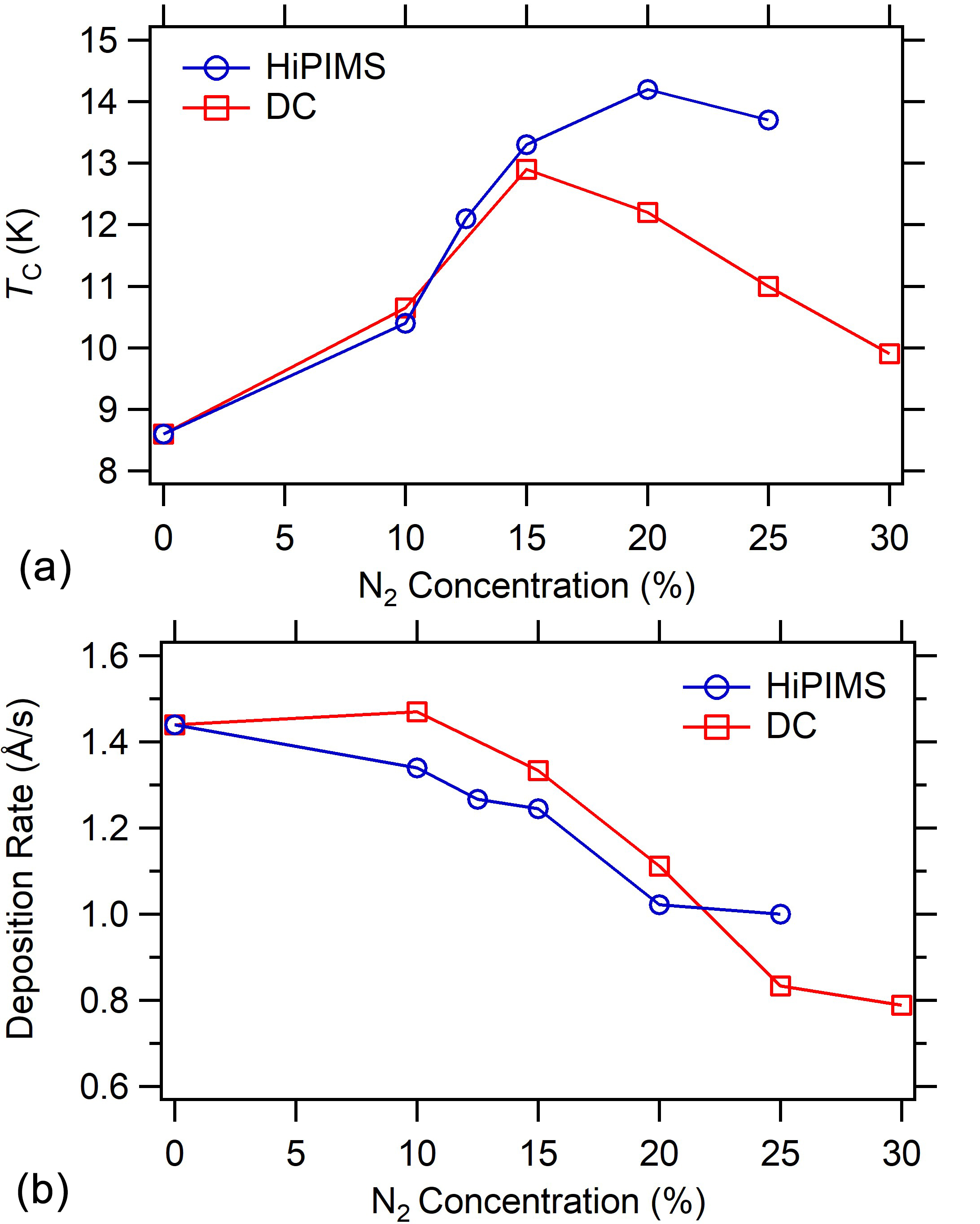}
\caption{\label{fig:VaryN2} (a) $T_\mathrm{C}$ of $120$ nm thick NbN films deposited via HiPIMS and DC sputtering onto unheated oxidized Si wafers with different nitrogen gas concentrations. (b) Deposition rate as a function of nitrogen gas concentration. DC sputtering was performed at $150$ W and HiPIMS was performed at a time-average power of $250$ W.}
\end{figure}

The resulting deposition rates are shown in figure \ref{fig:VaryN2}b. After these were determined, $120$ nm thick films were deposited at different nitrogen gas concentrations using both DC sputtering and HiPIMS, and the $T_\mathrm{C}$ of each film was measured. Here we define $T_\mathrm{C}$ as the temperature at which the resistance drops to half of its value at $18$ K. Figure \ref{fig:VaryN2}a shows these results. The nitrogen concentration yielding the highest $T_\mathrm{C}$ was 15\% for DC sputtering and 20\% for HiPIMS. The higher nitrogen gas concentration for the optimal HiPIMS film suggests that the HiPIMS process enables us to get closer to $1{:}1$ stoichiometry before beginning to form the $\epsilon$-NbN crystal phase that reduces $T_\mathrm{C}$.

\begin{figure}
\includegraphics[width=8.5cm]{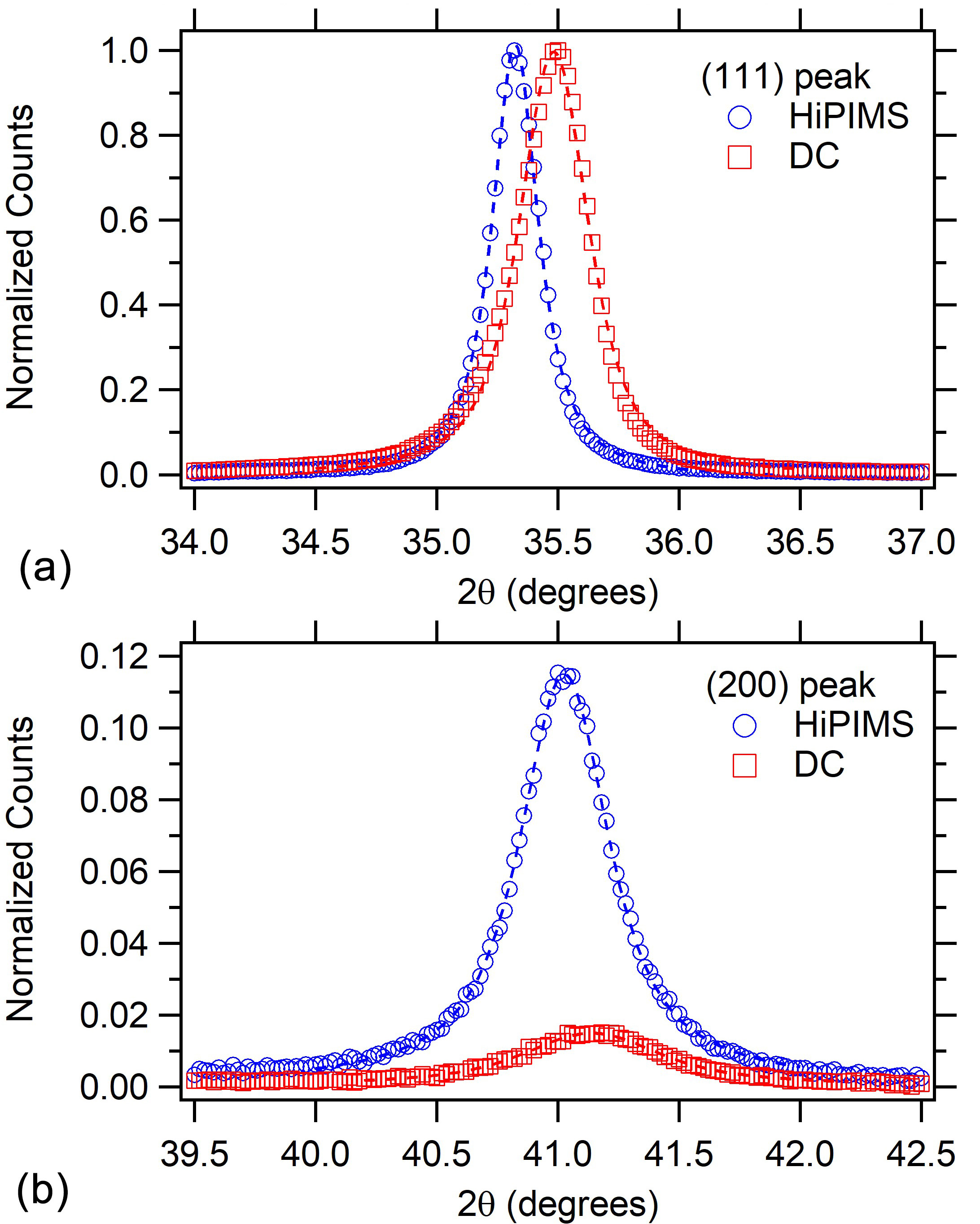}
\caption{\label{fig:XRD} X-ray diffraction spectrum showing measured peaks (markers) along with pseudo-Voigt fits (dashed lines). Samples are $1$ $\mu$m thick films deposited via HiPIMS and DC sputtering using the optimal nitrogen concentration for each process. (a) shows the (111) peaks and (b) shows the (200) peaks, and and each spectrum has been normalized to the height of the (111) peak.}
\end{figure}

To gain further insight into the film composition and microsctructure, x-ray diffraction (XRD) $\theta$-$2\theta$ scans were performed with a Shimadzu XRD-6100 on films deposited via DC sputtering and HiPIMS at the optimal nitrogen concentration for each process. To increase the XRD signal, thicker films of $1$ \textmu m were grown for this characterization. $2\theta$ peaks were observed at approximately $35^\circ$ and $41^\circ$, as shown in figure \ref{fig:XRD}. Each measured spectrum has been normalized to the height of the larger peak. The observed peaks are consistent with Miller indices ($hkl$) = (111) and (200) for a cubic crystal structure. For a simple cubic structure, the lattice constant $a$ is related to the interplane spacing $d$ by $a = d\sqrt{h^2 + k^2 + l^2}$. The diffraction angle $\theta$ is related to $d$ via Bragg's law, $n\lambda = 2d\sin\theta$, with $n=1$. The XRD system uses a Cu K$_\alpha$ x-ray source with a wavelength $\lambda = 1.5418$ {\AA}.\cite{XRDtutorial} Each measured peak was fit to a pseudo-Voigt function, the weighted sum of a Gaussian and a Lorentzian. The peak center from the fit was then used to calculate the lattice constant $a$ using the specified Miller indices. The results are summarized in table \ref{tab:table1}. Also included in the table are the values of the full-width at half-maximum (FWHM) of each peak.    

The HiPIMS film yielded a lattice constant of $4.400$-$4.401$ {\AA} and the DC sputtered film yielded a lattice constant of $4.382$-$4.388$ {\AA}. Previous work has shown that an increase in nitrogen concentration is associated with an increase in the lattice constant.\cite{shy,kalal,torche} The larger lattice constant in the \mbox{HiPIMS} film thus confirms that this film has higher nitrogen content than the DC sputtered film, consistent with the picture of the HiPIMS process enabling us to get closer to optimal stoichiometry before beginning to form the $\epsilon$-NbN crystal phase that reduces $T_\mathrm{C}$.

\begin{table}
\caption{\label{tab:table1}Cubic lattice constant values calculated from XRD data in figure \ref{fig:XRD} assuming a simple cubic crystal structure and assigning Miller indices of (111) to the peak at $35^{\circ}$ and Miller indices of (200) to the peak at $41^{\circ}$.}
\begin{ruledtabular}
\begin{tabular}{cccccc}
\multicolumn{3}{c}{HiPIMS film} &\multicolumn{3}{c}{DC sputtered film}\\
\begin{minipage}{0.15\linewidth}
$2\theta$\\peak\end{minipage}
&\begin{minipage}{0.15\linewidth}FHWM\end{minipage}
&\begin{minipage}{0.15\linewidth}lattice\\constant\end{minipage}
&\begin{minipage}{0.15\linewidth}$2\theta$\\peak\end{minipage}
&\begin{minipage}{0.15\linewidth}FWHM\end{minipage}
&\begin{minipage}{0.15\linewidth}lattice\\constant\end{minipage}
\\[2mm]   
\hline
$35.324^{\circ}$ & $0.237^{\circ}$ & $4.401$ {\AA} & $35.483^{\circ}$ & $0.341^{\circ}$ & $4.382$ {\AA}
\\
$41.027^{\circ}$ & $0.441^{\circ}$ & $4.400$ {\AA} & $41.141^{\circ}$ & $0.716^{\circ}$ & $4.388$ {\AA}
\\
\end{tabular}
\end{ruledtabular}
\end{table}

The peak width contains contributions from crystallite size, film disorder (including strain), and instrument resolution. In the pseudo-Voigt fit, the width of the Lorenzian component is attributed to crystallite size, while the width of the Gaussian component is attributed to nonuniform strain.\cite{XRDtutorial} For all our peak fits, the Lorentzian component dominated -- the pseudo-Voigt fit was nearly identical to a Lorentzian fit -- suggesting that the dominant source of intrinsic peak broadening is crystallite size. Assuming that the sample broadening is due only to finite crystallite size, we can determine the volume-average crystallite size $L_V$ using the Scherrer formula, $L_V = K\lambda/(\beta_{s}\cos\theta)$, where $K$ is the unitless structure factor, which for simplicity we assume to be $0.94$ for spherical crystallites, $\beta_s$ is the intrinsic sample broadening (in radians), and $\theta$ is the peak angle. The sample broadening is found from $\beta_{s} = \sqrt{\beta_{m}^2-\beta_{i}^2}$, where $\beta_{m}$ is the measured FWHM of each peak and $\beta_i$ is the contribution from the instrument resolution. The instrument resolution was quantified by measuring a Si powder reference, whose (111) peak was centered at $2\theta = 28.406^\circ$ with a FWHM of $\beta_{i} = 0.129^\circ$. Using this approach, for the (111) peak of the HiPIMS film we get $L_V = 48$ nm, and for the (111) peak of the DC sputtered film we get $L_V = 29$ nm. We note that the (200) peaks are broader and yield smaller crystallite sizes than the (111) peaks, although the trend of the HiPIMS film having a larger crystallite size still holds. This discrepancy could be related to asymmetrical crystallites, as each peak width reflects the crystallite size in that particular (hkl) direction, and it could also be related to other sources of peak broadening not captured by the Scherrer equation. Nonetheless, the narrower peak widths in the XRD spectrum of the HiPIMS film indicate that the HiPIMS process produces films with a higher degree of crystallinity than DC sputtering.

%While there are more sophisticated approaches to analyzing peak widths that enable one to determine different contributions to the broadening, these approaches generally require more than two peaks, ideally distributed over a reasonably large range of angles, to yield reliable results.[ref] Nonetheless, the larger crystallite size for the HiPIMS film is consistent with the picture of the HiPIMS process promoting a higher degree of film crystallinity.   

\begin{figure}
\includegraphics[width=8.5cm]{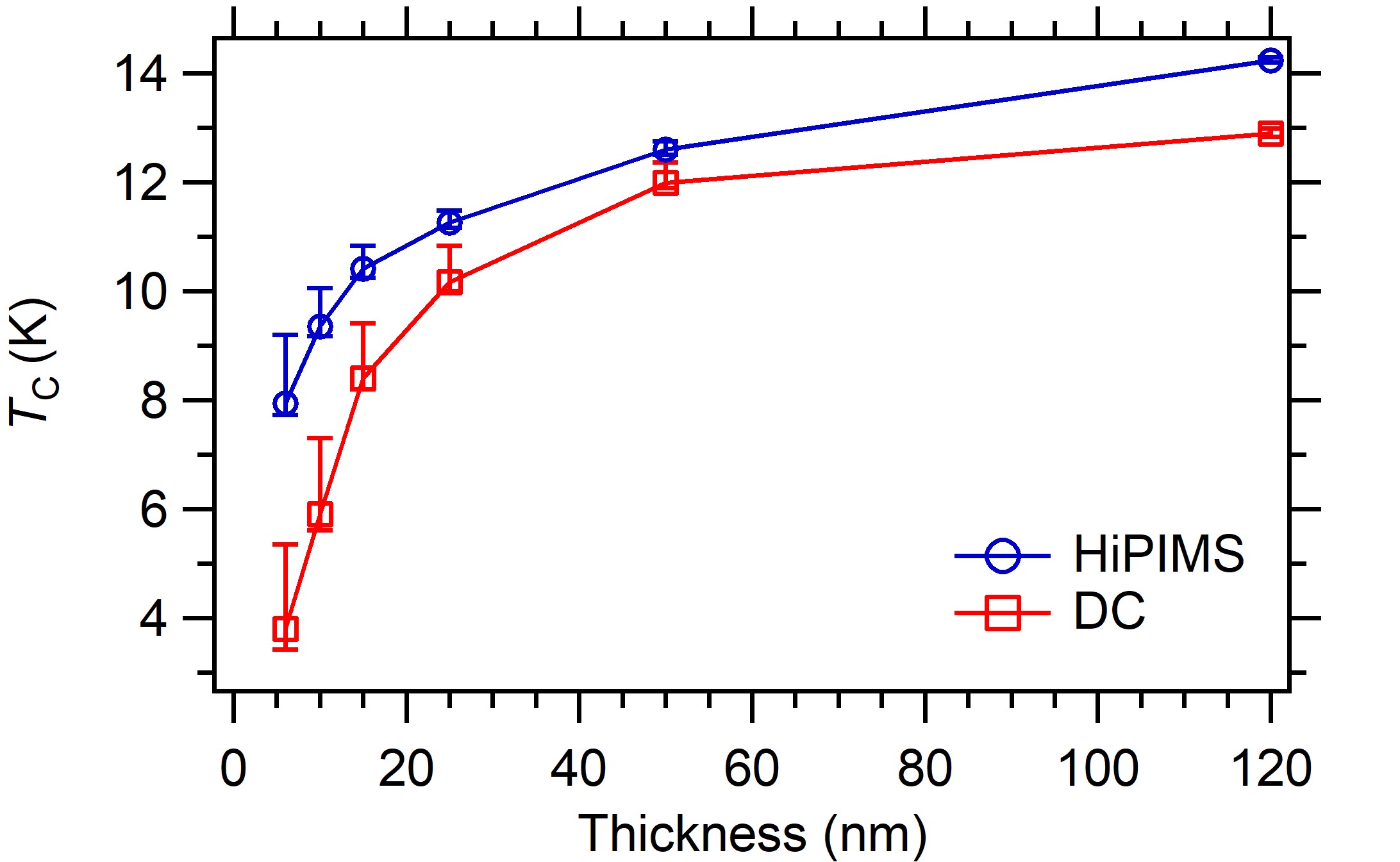}
\caption{\label{fig:Thickness1} $T_\mathrm{C}$ of NbN films deposited via HiPIMS and DC sputtering as a function of film thickness. Error bars show the width of the transition, defined as 90\% to 10\% of the normal-state resistance at $18$ K. Films were deposited onto unheated Si substrates with a native oxide using the optimal nitrogen concentration for each deposition technique.}
\end{figure}

\begin{figure}
\includegraphics[width=8.5cm]{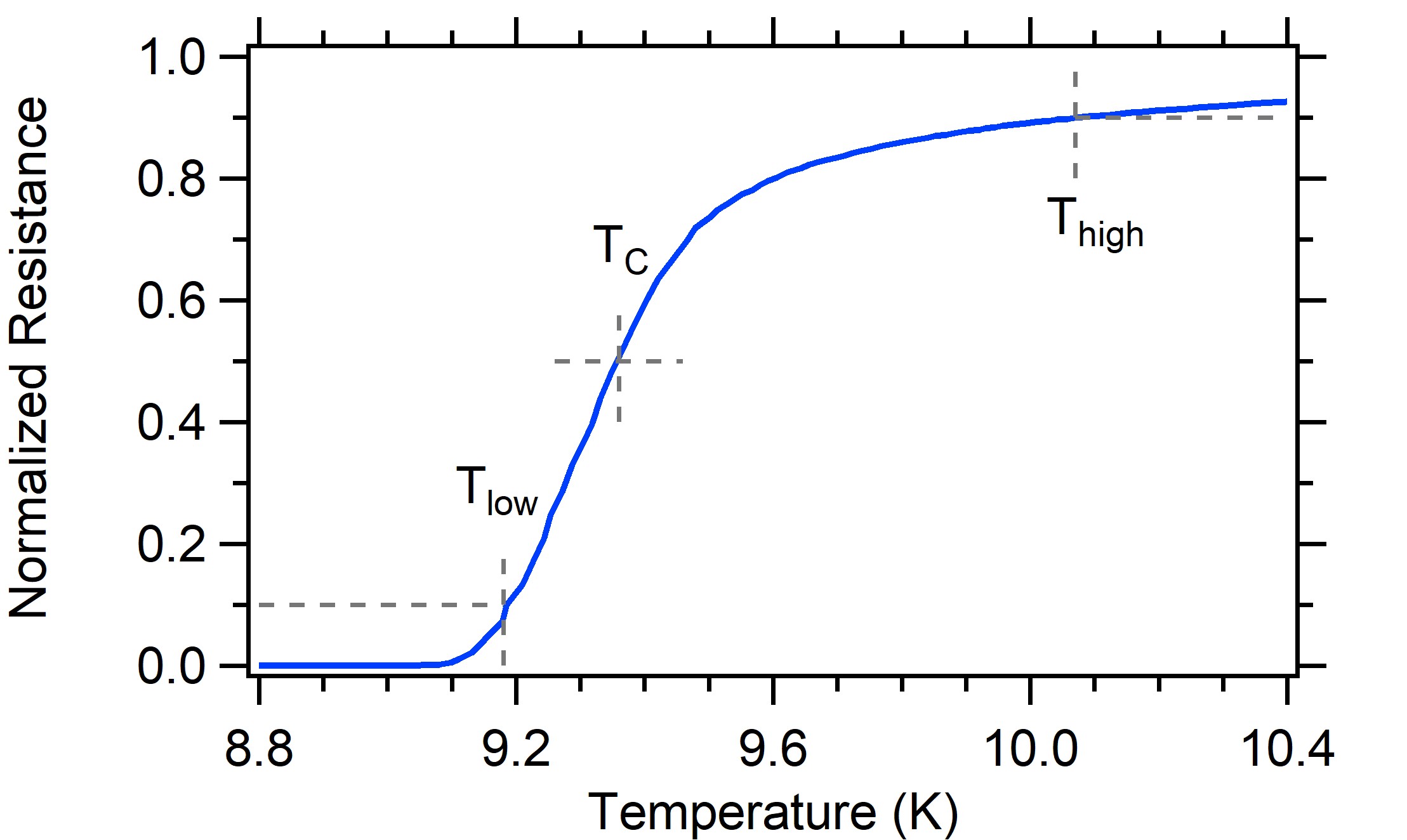}
\caption{\label{fig:RT} Resistance versus temperature curve from a $10$ nm thick NbN film deposited via HiPIMS with $20\%$ nitrogen gas concentration onto an unheated substrate. The resistance has been normalized to the value at $18$ K, and $120$ $\Omega$ of lead resistance was subtracted before normalization. $T_\mathrm{C}$ is defined as the temperature at which the resistance is half the value at $18$ K. The transition width $\Delta T_\mathrm{C} = T_{high} - T_{low}$, where $T_{high}$ is the temperature at which the resistance is $90\%$ of the value at $18$ K and $T_{low}$ is the temperature at which the resistance is $10\%$ of the value at $18$ K.}
\end{figure}

Using the optimal nitrogen concentrations for the DC and HiPIMS processes, a set of films of varying thickness between $6$ nm and $120$ nm were deposited in order to explore the dependence of $T_\mathrm{C}$ on the film thickness. The results are shown in figure \ref{fig:Thickness1}. As before, $T_\mathrm{C}$ is defined as the temperature at which the resistance drops to half the value at $18$ K. We see that the HiPIMS films have a higher $T_\mathrm{C}$ than the DC sputtered films at all thicknesses studied, with the difference becoming larger for the thinnest samples. The error bars in figure \ref{fig:Thickness1} show the transition width $\Delta T_\mathrm{C} = T_{high} - T_{low}$, where $T_{high}$ is the temperature at which the resistance is $90\%$ of the value at $18$ K and $T_{low}$ is the temperature at which the resistance is $10\%$ of the value at $18$ K (after subtraction of the lead resistance). The transition width broadens as the film thickness is decreased and develops a noticeably asymmetry, with $(T_{high}-T_\mathrm{C}) > (T_\mathrm{C}-T_{low})$, as seen in the example resistance versus temperature curve shown in figure \ref{fig:RT}. This asymmetry is attributed to fluctuations of the superconducting order parameter above $T_\mathrm{C}$, which are enhanced in systems with low dimensionality and strong disorder.\cite{Larkin,Baeva}

\section{Optimization of HiPIMS Films}

\begin{figure}
\includegraphics[width=8.5cm]{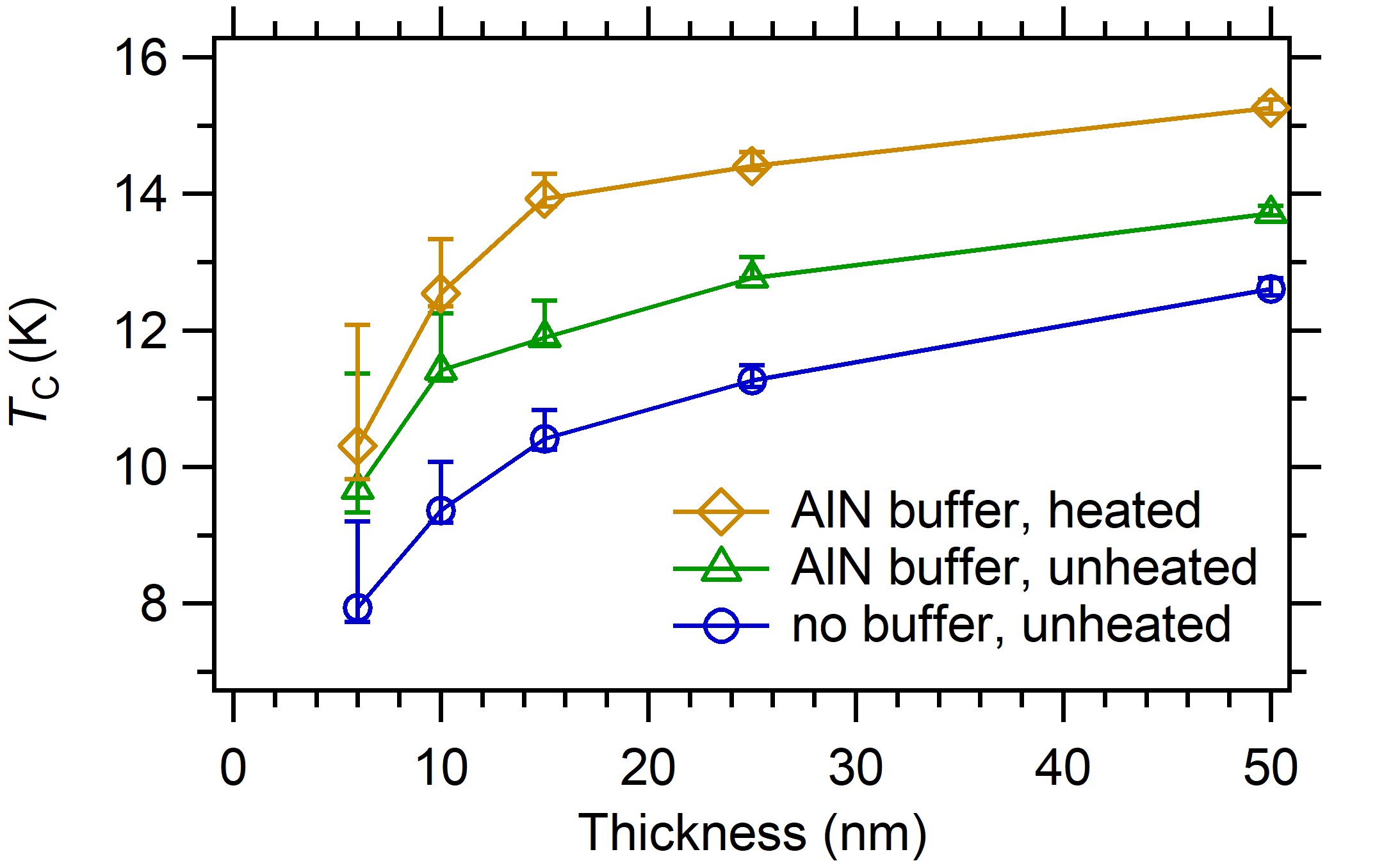}
\caption{\label{fig:Thickness2} $T_\mathrm{C}$ as a function of thickness of NbN films deposited via HiPIMS with 20\% nitrogen concentration. Deposition onto unheated Si substrates with a native oxide is compared to films deposited on the same substrate with the addition of a $20$ nm thick AlN buffer layer, as well as films with a $40$ nm thick AlN buffer layer that were heated to $350$ $^{\circ}$C for the NbN deposition. Error bars show the width of the transition, defined as 90\% to 10\% of the normal-state resistance at $18$ K.}
\end{figure}

To study the potential for improvement of the $T_\mathrm{C}$ of our HiPIMS films, we deposited a set of films of different thicknesses on top of a $20$ nm AlN buffer layer, and on top of a heated substrate with a $40$ nm AlN buffer layer. The AlN layer was deposited as described previously, and then the NbN was deposited in the same deposition system without breaking vacuum. All NbN films were deposited using HiPIMS with a $20$\% nitrogen gas concentration. The resulting critical temperatures are shown in figure \ref{fig:Thickness2}, with the error bars showing $\Delta T_\mathrm{C}$. As has been seen in previous work with DC sputtering, the use of the AlN buffer layer increased $T_\mathrm{C}$.\cite{shiino,rhazi2021improvement} This is attributed to lattice matching between the c-plane of hexagonal AlN and the (111) plane of cubic NbN.

We also investigated the effect of depositing onto heated substrates. Initially, the quartz lamp heater in the deposition system was powered with an AC power supply. This produced an instability in the HiPIMS process, with the plasma visibly flickering at the power line frequency. Powering the heater with a DC power supply eliminated this instability. All heated depositions presented here used the DC power supply. 

First, we deposited NbN onto bare substrates (Si with a native oxide and no AlN buffer layer) that were heated to $350$ $^\circ$C. The resistance versus temperature curves of the resulting films showed a reduced and significantly broadened superconducting transition ($\Delta T_\mathrm{C} > 1$ K for relatively thick films). Films deposited with DC sputtering onto substrates heated to the same temperature did not show this degradation of $T_\mathrm{C}$. Thus we believe that the combination of thermal energy and the high kinetic energy produced by the HiPIMS pulses enables an unwanted chemical interaction between the substrate and the NbN. We found that this could be prevented, and relatively sharp transitions could be obtained, by first depositing the AlN layer at room temperature, then heating the substrate to $350$ $^\circ$C, and then depositing the NbN. For these samples, the thickness of the AlN buffer layer was increased to $40$ nm to ensure that it was effective at inhibiting the interaction between the NbN and the underlying substrate. With the buffer layer, the heated NbN films showed a further increase in $T_\mathrm{C}$, as seen in figure \ref{fig:Thickness2}. This is consistent with the expectation that a higher substrate temperature results in greater adatom mobility during the deposition process, which improves crystallization as local nitrogen excesses and deficits in the lattice can diffuse out and long-range crystalline order can develop.\cite{kalal}

We also characterized the effect of different deposition parameters on the normal-state resistivity of the films. At room temperature, the Si ($\rho > 5$ k$\Omega$ cm) with a native oxide layer creates a finite-resistance pathway in parallel with the film. This can be avoided by using a thicker oxide layer. To ensure accurate resistivity measurements at all temperatures, we deposited a set of $10$ nm thick NbN films on Si substrates with a $280$ nm thick thermal oxide (SiO$_2$). We verified that the resulting films had the same $T_\mathrm{C}$ as those deposited on Si substrates with a native oxide. Films were deposited using both DC sputtering and HiPIMS on unheated substrates, on unheated substrates with a $20$ nm thick AlN buffer layer, and on substrates with a $40$ nm AlN buffer that were then heated to $350$ $^\circ$C. The nitrogen gas concentration was $15$\% for DC sputtering and $20$\% for HiPIMS. The resulting sheet resistance values at room temperature and at $18$ K are given in table \ref{tab:table2.2}, along with the $T_\mathrm{C}$ and transition width $\Delta T_\mathrm{C}$. As the resistivity of very thin NbN films tends to increase over time due to oxidation, care was taken to measure each film promptly after deposition. 

\begin{table}
\setlength\extrarowheight{7pt}
\caption{\label{tab:table2.2}Comparison of 10 nm thick NbN films}
\begin{ruledtabular}
    \begin{tabular}{ccccc}
        \begin{minipage}{0.08\textwidth}Deposition\\ type\end{minipage}&\begin{minipage}{0.08\textwidth}Sheet resistance at 296 K\end{minipage}&\begin{minipage}{0.08\textwidth}Sheet resistance at 18 K\end{minipage}&\begin{minipage}{0.08\textwidth}$T_\mathrm{C}$\end{minipage}&\begin{minipage}{0.08\textwidth}$\Delta T_\mathrm{C}$\end{minipage}\\[4mm]
        \hline
        DC & $652$ $\Omega$/sq & $908$ $\Omega$/sq & $5.90$ K & $1.70$ K\\
        HiPIMS & $281$ $\Omega$/sq & $328$ $\Omega$/sq & $9.36$ K & $0.89$ K\\
        \begin{minipage}{0.115\textwidth}DC with AlN buffer\end{minipage} & 921 $\Omega$/sq & 1,613 $\Omega$/sq & 7.29 K & $2.72$ K\\ 
        \begin{minipage}{0.115\textwidth}HiPIMS with AlN buffer\end{minipage}& $346$ $\Omega$/sq & $419$ $\Omega$/sq & $11.42$ K & $0.97$ K\\
       \begin{minipage}{0.115\textwidth}heated DC with AlN buffer\end{minipage} & 691 $\Omega$/sq & 1,129 $\Omega$/sq & 8.96 K & $2.32$ K\\
        \begin{minipage}{0.115\textwidth}heated HiPIMS with AlN buffer\end{minipage}& $323$ $\Omega$/sq & 391 $\Omega$/sq & $12.44$ K & $0.98$ K\\        
    \end{tabular}
    \end{ruledtabular}
\end{table}

In each case, the HiPIMS process produced films with a lower sheet resistance and higher $T_\mathrm{C}$ than DC sputtering. Interestingly, and in contrast to previous results,\cite{shiino} we observe an increase in the sheet resistance when the film is deposited on an AlN buffer layer as compared to the bare substrate. To determine if this is related to the surface roughness of the AlN, a $20$ nm thick AlN film was characterized via atomic force microscopy. The resulting root mean square roughness of $0.4$ nm is not likely to significantly affect the resistivity of $10$ nm thick NbN films. It may be that the increased resistivity is related to film strain, although further studies would be required to test this hypothesis.

\section{Conclusions}

In reactive DC sputtered NbN films, it has been shown that a hexagonal crystal phase with a lower $T_\mathrm{C}$ begins to form as one approaches a $1{:}1$ stoichiometric ratio.\cite{benkahoul2004structural,wright} We have found that, for reactive HiPIMS, the nitrogen gas concentration that produces the highest $T_\mathrm{C}$ is greater than that for DC sputtering, and the maximum  $T_\mathrm{C}$ for the HiPIMS films is greater than the maximum $T_\mathrm{C}$ for the DC sputtered films. This holds true across all film thicknesses studied, from $6$ nm to $120$ nm. We believe that this is the result of the HiPIMS process enabling the films to get closer to optimal stoichiometry before beginning to form the hexagonal $\epsilon$-NbN crystal phase. The lower normal-state resistivity of the HiPIMS films is attributed to improved film crystallinity enabled by the higher kinetic energy in the HiPIMS process. These observations are consistent with XRD analysis, which show that the HiPIMS films have a larger lattice constant, consistent with greater nitrogen concentration, and a narrower peak width, consistent with a larger crystallite size. 

We found that the $T_\mathrm{C}$ of HiPIMS films can be increased through the use of an AlN buffer layer, and by depositing an AlN buffer layer and then subsequently heating the substrate to $350$ $^\circ$C. In all deposition scenarios tested, the HiPIMS films have a higher $T_\mathrm{C}$ and a lower normal-state resistivity than the DC sputtered films. 

The HiPIMS deposition process presents a large parameter space, and further exploration of this space may be productive. For example, we have not yet studied the effect of varying parameters such as the deposition chamber pressure and the height and duration of the HiPIMS voltage pulses, including the positive kick pulse. In future work, we also plan to explore the benefits of the HiPIMS process for other materials in the superconducting transition metal nitride family.

\begin{acknowledgments}
This work was supported by the National Science Foundation through awards ECCS-2117007 and ECCS-2000778. We thank Drs. T.J. Mullen and G.A. Wurtz for performing AFM characterization.
\end{acknowledgments}

\section*{Data Availability Statement}
The data that support the findings of this study are available from the corresponding author upon reasonable request.

%\nocite{*}
\bibliography{aipsamp}% Produces the bibliography via BibTeX.

\end{document}